# Compression Effects and Scene Details on the Source Camera Identification of Digital Videos

*Group of Analysis, Security and Systems (GASS), Department of Software Engineering and Artificial Intelligence (DISIA), Faculty of Computer Science and Engineering, Office 431, Universidad Complutense de Madrid (UCM), Calle Profesor José García Santesmases 9, Ciudad Universitaria, 28040 Madrid, Spain*

Raquel Ramos López, Ana Lucila Sandoval Orozco, Luis Javier García Villalba*

**Abstract**

The continuous growth of technologies like 4G or 5G has led to a massive use of mobile devices such as smartphones and tablets. This phenomenon, combined with the fact that people use mobile phones for a longer period of time, results in mobile phones becoming the main source of creation of visual information. However, its reliability as a true representation of reality cannot be taken for granted due to the constant increase in editing software. This makes it easier to alter original content without leaving a noticeable trace in the modification. Therefore, it is essential to introduce forensic analysis mechanisms to guarantee the authenticity or integrity of a certain digital video, particularly if it may be considered as evidence in legal proceedings. This paper explains the branch of multimedia forensic analysis that allows to determine the identification of the source of acquisition of a certain video by exploiting the unique traces left by the camera sensor of the mobile device in visual content. To do this, a technique that performs the identification of the source of acquisition of digital videos from mobile devices is presented. It involves 3 stages: 1) Extraction of the sensor fingerprint by applying the

---

*Corresponding author

*Email addresses:* `raqram01@ucm.es`, Securitas Direct. Madrid, Spain (e-mail: `raquel.rlopez@securitasdirect.es`) (Raquel Ramos López), `asandoval@fdi.ucm.es` (Ana Lucila Sandoval Orozco), `javiergv@fdi.ucm.es` (Luis Javier García Villalba)

block-based technique. 2) Filtering the strong component of the PRNU signal to improve the quality of the sensor fingerprint. 3) Classification of digital videos in an open scenario, that is, where the forensic analyst does not need to have access to the device that recorded the video to find out the origin of the video. The main contribution of the proposed technique eliminates the details of the scene to improve the PRNU fingerprint. It should be noted that these techniques are applied to digital images and not to digital videos. In this work, we show that it is necessary to take this improvement into account to improve the identification of digital videos. Experimental results are also presented that support the validity of the techniques used and show promising results.

## 1. Introduction

Mobile technology can be considered a success in the history of telecommunications due to the fact that is has become very popular and widespread in our lives. Throughout history, a constant cycle of evolution has been observed when it comes to a new generation of mobile network innovation that is launched every decade. As shown in Figure 1 below, since the introduction of 1G technology in the late 1970s, there has been a fairly regular pace of improvements in mobile capabilities over the past four decades, plus the gradual incorporation of elements that until then were only intended for personal computers (processor, qwerty keyboards, larger screens ...), together with the integration of specific software programs for these devices. All this has made mobile devices a feasible alternative to desktop computers. The introduction of each new generation has not only improved network performance, but has made newer and more advanced applications and devices available. Similarly, with 5G, we expect both network improvements in terms of bandwidth (1+ Gbps) and ultra-low latency, as well as features such as improved energy efficiency, cost optimization, massive IoT density, and dynamic resource allocation in order to enable a broad spectrum of wireless applications and IoT connections Cis:2022.

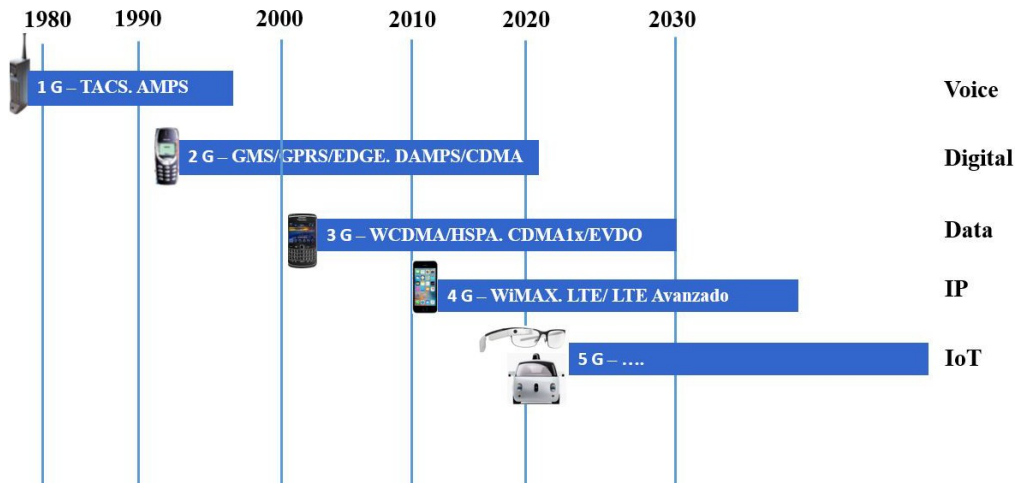

Figure 1: Evolution of the mobile communication network

Smartphones have become an indispensable companion in our day to day lives. At their most basic level, we use mobile devices to be connected with our friends and family and to access the Internet.

However, there is a functionality that is progressively gaining importance, and in which manufacturers are putting more and more effort: the use of the camcorder. The appearance of these devices has led to an important change in people's behavior and communication habits. In a very short time, these devices have become the focus of interest for a great majority of people, so that a large part of their daily activities are channeled through their smartphones. Currently, there are increasingly sophisticated video manipulation tools that make it difficult to identify the characteristics of interest. This means that what a video represents cannot be taken for granted. Therefore, it is critical to employ techniques and research that are capable of determining, unmistakably, if a video has been manipulated or, on the contrary, shows its original content. The forensic analysis of videos has proven to be more difficult compared to the analysis of images, since the data contained in the videos has higher compression formats, which can compromise the existing "fingerprints" and make it more difficult to recover the processing of a video from its origin.

One of the first works where the compression of a video was taken into account was in Houten and Geradts ([2009](#)), where the use of the PRNU footprint on videos coming from YouTube was analysed. To carry out their experiments, they used a set of webcams and codecs to record and encode videos. These videos were then uploaded to YouTube and downloaded. The results obtained were good but the codecs used have already become obsolete, so nowadays, this technique could not be used in a real scenario.

For the identification of the source of acquisition there are two main approaches: closed and open scenarios. A scenario is considered to be closed when the identification of the video source is carried out on a specific set of camcorders known a priori. For this approach, a set of videos from each camcorder is usually used to train a classifier and later predict the source of acquisition of the videos under investigation. One of the most used techniques for classifying digital videos is SVM (Support Vector Machine). In an open scenario, the forensic analyst does not know the set of video cameras to which the videos to be identified belong. In this type of classification, in which there is no data on videos known a priori, the objective is not to identify their brand and model, but rather to be able to group together the frames of the videos that belong to the same video.

In Meij and Geradts (2018) an attempt was made to determine the origin of a digital video from a mobile device that was shared by the WhatsApp messaging platform on both iOS and Android operating systems. They used videos from ten different camcorder models to carry out their experiments. For each of these cameras, three types of videos were recorded: one video was recorded in an indoor space, another one in an outdoor space, and another that recorded a gray surface to be used as a reference. The results indicated that it is possible to determine the source camera of a digital video with a high success rate, because the videos that come from the same camera show many similarities. Once these videos have been transmitted through WhatsApp, this precision decreases, although it is still possible to classify the vast majority of them with somewhat worse results in the case of iOS. This method is very limited to be applied in open scenarios for two reasons. On the one hand, the success rate decreases in iOS devices, and on the other hand, this method does not take into account the wide variety of platforms through which a video can be distributed, like in the case of video sharing platforms such as YouTube.

It should be noted that in the scientific literature, the contributions that exist today focus on improving the sensor footprint applied to the study of digital images. However, this study is almost non-existent in the case of digital videos. This is a gap that we aim to bridge with this work, given that it is possible to improve the PRNU fingerprint of the sensor in the case of digital videos.

One of the initial works applied to digital images, where an improvement in the quality of the PRNU fingerprint of the noise was proposed, was in Li (2010). Their idea was based on the following hypothesis: The details of the scene when representing high-frequency components make their magnitude much greater than that of the noise pattern. For that reason, it is necessary to eliminate the fragments of the high-frequency scene to improve the sensor noise footprint. Therefore, in this work, an approach is proposed to mitigate the influence of scene details when calculating the noise pattern, and thus improve the success rate when identifying the origin of the device.

Another work that focuses on improving the PRNU footprint applied to images is in Akshatha et al. (2016). After estimating the PRNU noise pattern of the images, they are represented according to the characteristics that can be used in the classification. The two sets of features are identified and extracted from the images. The first set includes Higher order wavelet statistics (HOWS) from the estimated PRNU noise. The second set consists of statistical features from the original images. The functions are combined and used to discriminate images based on their source cameras. To carry out their experiments, they used 4 different groups of two cameras each, and they obtained an average hit rate of 100%, 99.75%, 94.75% respectively for each of the groups. As we have already mentioned, the technique is based on digital images but does not take into account digital videos. Another limitation of this technique is that it has not been tested when an image is shared by WhatsApp or downloaded from the YouTube platform.

The techniques using the PRNU sensor fingerprint to carry out identification processes need to analyze the content of the video. They have become robust and reliable techniques, yet they require a higher computational cost. There are other methods such as the one presented in López et al. (2020) where they identify the source of digital video acquisition by analyzing the elemental structure of a video called the atom. An atom is made up of a set of labels, which are values in a hierarchical way. The value of each of these atoms gives some clues to obtain the origin of a mobile device. For a more in-depth study of the container atoms, see Gloe et al. (2014). To perform the classification, they used a hierarchical grouping and another one based on density (OPTICS). With the hierarchical grouping, they achieved a coefficient of "homogeneity" higher than 0.9% when it came to identifying the brand and 0.80% when it came to identifying the model. As for the OPTICS density-based algorithm, they achieved a homogeneity of 0.97% grouping by brand and 0.87% grouping by model. This technique is much faster than techniques that have to go through video content to find out the origin of the device.

In this work, it was necessary to use a smaller set of videos instead of the one proposed in López et al. (2020) due to the high computational cost required by the technique that processes the video content instead of the atoms or video labels to identify the video source.

In the field of engineering, computer simulations are widely used to replace experiments with physical models because these simulations are often computationally expensive. However, many model-based engineering design problems require numerous simulations to reach an acceptable solution. This can be computationally prohibitive. Often a single surrogate model or meta-model is used to replace a detailed simulation in design problems which require repeated calculations. This surrogate is obtained using information derived from the physical model Alizadeh et al. (2019).

In Alizadeh et al. (2019) proposed a method based on cross-validation to find an ensemble of surrogates(EoS) which is created by the least possible number of data points. The resulting ensemble surrogate has higher accuracy than each individual surrogate and is less computationally intensive. To achieve this ensemble surrogate, we compare it with individual surrogate models based on three main factors: the size of the problem, precision and calculation time. They found that it is effective to use cross validation. They obtained the highest accuracy level with least required data and less computation time by using the right number of samples. An example of surrogates is relatively insensitive to the size of the sample data or number of data points.

In Alizadeh et al. (2020) they created a practical guidance based on a trade-off among three main drivers, namely, size (how much information is necessary to compute the surrogate model), accuracy (how accurate the surrogate model must be) and computational time (how much time is required for the surrogate modeling process). To make their proposal, they reviewed the latest generation surrogate models on more than 200 articles from the most recent literature.

The main contribution of this work is threefold:

1. A digital video acquisition source identification technique is proposed. It takes into account the effect produced by video compression to obtain the sensor noise fingerprint, by applying the block-based technique proposed in Kouokam and Dirik (2019).

2. Once the blocks that have survived compression have been calculated, the method proposed in Li (2010) is used. It improves the PRNU noise pattern extracted from the videos under investigation. Therefore, in this work it is shown that in order to have a reliable and robust hit rate, it is not enough to review the effect that compression has on the videos, but also techniques that improve the fingerprint formed by the imperfections of the videosensor. Related works can be found in the literature that improve the sensor footprint applied to digital images but not to digital videos.

3. Finally, a grouping algorithm that classifies the videos in an open setting is used. In this case, the forensic analyst does not have access to the specific mobile device where the criminal activity is located, since you can find the video under investigation on the internet. Furthermore, it is impossible for the analyst to have access to a video from the vast variety of mobile devices available on the market.

The remainder of the paper is structured as follows. Section 2 provides the reader with some background on algorithms for identifying the source of digital videos and provides the formal definition of the problem. Section 3 reports all technical details about the proposed solution. Section 4 presents the numerical experiments carried out to validate the proposed method while Section 5 concludes the paper providing some conclusive remarks.

## 2. Digital Video Source Identification Techniques

This section details the main techniques for forensic analysis of digital videos, with an emphasis on the techniques for identifying the source of the video, as this is the branch of forensic analysis on which this work is focused. The section after next discusses the most important techniques related to video content analysis, with a special emphasis on the methods that study the sensor's noise pattern.

According to Sandoval Orozco et al. (2013), there are five main groups: 1) metadata, 2) matrix defects CFA, 3) image features, 4) interpolation color, 5) sensor imperfections combined with Wavelet transformations.

The techniques based on sensor imperfections have gradually evolved into two different families: one based on pixel defects and another that analyzes the noise pattern in the sensor. The latter is the most used technique for identifying sources in both images and digital videos. This is the basis of this work, where the demonstrations of Lukas et al. (2006) have been taken into account. It was ultimately determined that the cameras generate a pattern of noise or SPN that can be used as a single classification method.

## 2.1. Analysis of the Sensor Noise Pattern

There are various sources of imperfections and noise introduced in the different stages of the digital video generation process. Even in the case of a uniform and fully lit photograph, small changes in pixel intensity can be observed. This is due to the shooting noise, which is random and largely due to the noise pattern, and is deterministic and stays roughly the same if multiple photos of the same scene are captured. The noise pattern in an image refers to any spatial pattern that does not change from one image to another and is composed of the spatial noise that is independent from the signal or fixed pattern noise FPN and the spatial noise due to the difference in response of each pixel to the incident signal or non-uniform response noise PRNU. The structure of the noise pattern is illustrated in Figure 2.

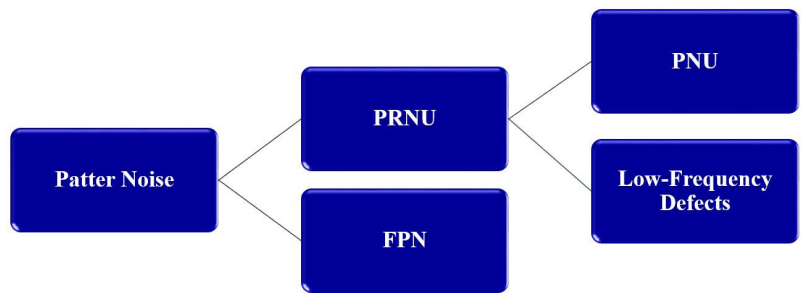

Figure 2: Sensor pattern noise.

The noise FPN is generated by the darkness current and is also dependent on exposure and temperature. Because a fixed pattern noise is an additive independent noise, some cameras automatically eliminate it by subtracting a dark frame from the images they produce. Noise PRNU is the dominant part of the noise pattern in images and is a multiplicative dependent noise that is mainly formed by pixel uniformity PNU and by low frequency defects such as zoom settings and the refraction of light in dust particles and lenses. The noise PNU is the difference in light sensitivity among the pixels in the sensor array. It is generated by the lack of homogeneity of the silicon wafers and imperfections during the manufacturing of sensors. Due to their nature and origin, it is highly unlikely that even sensors from the same wafer show correlated PNU patterns. This noise is not affected by room temperature or humidity. Noise PNU is typically more common, complex and significant in CMOS type of sensors due to the complexity of the pixel array circuitry Sandoval Orozco et al. (2015).

The noise pattern PRNU is produced by the variation of sensitivity to light of the individual pixels, due to the lack of homogeneity and impurities in the silicon chips, and to the imperfections resulting from the manufacturing of the sensor. In the case of videos, it may seem that estimating the pattern PRNU of a video camera from a video sequence is simpler than in the case of still images, due to the large number of frames available in a video. However, this is not true for two main reasons; first, the spatial resolution of videos is much lower than the spatial resolution of still images. Second, video frames generally have higher compression ratios than images compressed in JPEG format.

One of the first works in which a camcorder fingerprint was used to identify digital images was in Kurosawa et al. (1999). They proposed a method called CCD Fingerprint to identify a camcorder from videotaped images. They recommended using defective pixels and the dark current of CCD chips for camcorder identification. This approach is limited because thermal noise can only be removed with dark frames, and the dark current property is a weak signal that does not survive video compression.

Over time it has been shown that the technique developed in Fridrich et al. (2006) that identifies digital images using non-uniform response noise PRNU provides a much more robust and reliable fingerprint.

In Chen et al. (2007) it was determined that it is not possible to use a single frame of the video to identify its origin since the spatial resolution of the video is much lower than in the case of still images, and each frame is subjected to complex compression systems (MPEG- X, H26X and variants). Therefore, they demonstrated that by taking advantage of the temporal resolution of the videos, in cases of low resolution (264 × 352 pixels) and with only a 10 minutes video clip, it was possible to identify the source of the video device. The experiments were performed with 25 video cameras and showed that just 40 seconds of video is enough to have reliable results. If the video quality is decreased (the compression ratio is increased) and the spatial resolution is decreased, it is necessary to increase the video clip time in order to obtain reliable results. With videos in Internet LP format and a resolution of 264 × 352 and 150 kb/sec, good results are obtained for video clips with a duration of 10 minutes. The experiments carried out are limited because they do not take into account videos from mobile devices.

In Chuang et al. (2011) the impact of compression on the identification of the source camera is explored using the non-uniformity of response to the photo PRNU extracted from compressed videos generated by mobile devices. They concluded that type I frames are more trustworthy (reliable) than type P frames when giving an accurate PRNU estimate. They also, however, observed that the hit rate increases considerably when using all the frames that make up a video. However, they considered that using all the frames of a video has a high computational complexity. Motivated by this observation and by the fact that type I frames of a video have different reliability than type P frames, they proposed a mechanism to reorder and assign weights to the frames of a video by assigning weights 2 : 1 to type I and P frames respectively. In this way, they showed that the lower the number of frames plus a suitable weight assignment the better the estimation PRNU. This technique definitely takes into account the effect of video compression when obtaining the sensor fingerprint. Although the proposed method takes into account how video compression affects the calculation of the sensor imperfections footprint, later techniques such as the one proposed by Kouokam and Dirik (2019) show that it is necessary to use all types of frames (I, B, P) that a video has to improve the hit rate.

In García Villalba et al. (2016) a digital video source identification scheme based on PRNU noise and SVM classifier was proposed. The classification of the videos was carried out by selecting those frames with a significant scene change using the color histogram function. A total of 81 features, which are the Wavelet components of the sensor, are used to train the SVM classifier with training videos. A total of 5 different devices from 5 different brands were used to train the SVM classifier. The results obtained show a success rate of 87% or 90%, depending on the video resolution. Since this technique uses a classifier SVM, it has the following limitation: you need to have a video of the same brand and model of the video you want to investigate, and this is almost impossible nowadays due to the wide variety of models available in the market. Additionally, in the extracted frames, this technique does not take into account the effect that compression produces when improving the classification.

In Altinisik et al. (2018), a procedure is described to extract the fingerprint that produces the sensor noise PRNU in non-stabilized videos, taking into account the H.264 video compression standard. This technique has two main goals. On the one hand, this technique tries to eliminate the filtering procedure that is applied in the H.264 compression process. On the other hand, it also tries to select the blocks that best estimate the noise pattern PRNU. This brings us to the conclusion that the macroblocks that are encoded using intra-frame prediction and the loop filter should both be avoided so as not to weaken the noise PRNU. Quite the opposite occurs in type B or P macroblocks, where the encoder performs the prediction assuming filtered blocks. Therefore, it is necessary to remove the loop filter on the decoder as it will not produce a successful rebuild. To compensate for this behavior, the decoding process must be modified to rebuild both a filtered and an unfiltered version of each macroblock. Filtered macroblocks should be used to reconstruct future macroblocks, and unfiltered ones should be used for fingerprint extraction. The experiments were performed with videos converted from 550 photographic images with various levels of compression.

In Kouokam and Dirik (2019) they identified the source of videos coming from mobile devices and downloaded from YouTube by using PRNU and by studying the effect that the compression of the videos produces when estimating the PRNU fingerprint of a digital video. For each frame of the video, they looked for the blocks in which the PRNU noise had not been completely degraded due to the effects of the compression of the video. They proved that the PRNU noise pattern in a block survives compression if the DCT-AC coefficients of the prediction residue of the block are not all zero. Therefore, to estimate the noise of PRNU video, only those blocks that have at least a non-zero DCT-AC coefficient in frames I, P and B are used, since they still have, at least partially, a certain amount of noise of PRNU, which makes the block valid to identify the source of acquisition in digital videos coming from mobile devices and downloaded from YouTube.

All the techniques seen in this section only take into account the effect produced by the compression of a video to calculate the PRNU footprint. However, they do not take into consideration that it is necessary to improve said PRNU footprint in order to improve the classification. This work not only takes into account the problem that video compression causes when calculating the fingerprint produced by sensor imperfections, but also those techniques that improve the sensor fingerprint produced by the details of the scene. The most relevant techniques that show how to improve the fingerprint of the sensor applied to digital images are detailed below.

Sensor imperfections are due to the lack of uniformity in the photographic response (PRNU), which contains important information about the sensor in terms of frequency content. This information makes it appropriate for different forensic applications of videos and digital images. The main inconvenience of existing methods for the PRNU extraction is the fact that the extracted PRNU fingerprint has fine details of the image, such as high frequency details (textures, edges). To fix this problem, it is necessary to apply some techniques to remove those details from the scenes in order to improve the quality of the PRNU fingerprint.

As mentioned in the introduction section, one of the initial works in which the technique for the improvement of the sensor fingerprint was used was in Li (2010). It should be noted that this technique is based on the suppression of the content of high-frequency details from the scene. Following this hypothesis, Li (2010), developed five models and two of them yielded the best results applied to digital images. Later it was found that Lis improvement model Li (2010) also suppresses useful PRNU components Kang et al. (2011).

In Gupta and Tiwari (2018), a pre-processing step is applied to PRNU extraction filters widely accepted within the scientific community for low-frequency and high-frequency components of the image separately. The best results are obtained when using the Mihcak filter. They call their technique pMihack filter, this pMihcak filter contains the least amount of high-frequency details of the image. As seen in this section, in the case of digital videos, there are techniques which try to identify the origin of a video by applying the technique of PRNU sensor imperfections and how this affects video compression. On the other hand, in the literature we find that in the case of images, the PRNU fingerprint obtained is not sufficient if an improvement to that fingerprint is not applied. Therefore, in this study we suggest a technique that takes into account both approaches to identify the source of a digital video.

The classification algorithm used in this work, combines a hierarchical clustering and a flat clustering for the separation of the groups. The use of the silhouette coefficient for the validation of the results has proved to yield good results since high TPRs were obtained. To compare the convergence rate of this algorithm with the rest of algorithms in the literature, you need to consider the references below: Giri and Bardhan (2014), Dubey et al. (2015), Tsao (2015), Yin et al. (2016), Kazemi et al. (2018), Duan et al. (2018), Hoseini S. et al. (2019), Gharaei et al. (2019a), Shah et al. (2020), Rabbani et al. (2020), Sarkar and Giri (2020), Giri and Masanta (2020), Sayyadi and Awasthi (2020), Shah et al. (2020), Gharaei et al. (2020), Awasthi and Omrani (2019), Gharaei et al. (2019b)

## 3. Technique Description

To address the issue of identifying the source of a video, the most promising methods take advantage of the unique noise traces that the camera sensors produce in the acquired videos. As observed in the previous section, to identify the source of a digital video produced by a mobile device, most of the proposed techniques only take into account the effects that compression produces in the calculation of the noise fingerprint of the said video Chen et al. (2007), Houten and Geradts (2009), Chuang et al. (2011) Altinisik et al. (2018).

The main contribution of this work is to identify the source of digital video acquisition, which focuses on the impact of video compression to obtain the trace left by the sensor noise pattern. Conversely, the proposal takes into account the fact that the magnitude of the details of the scene that a video has tends to be much greater than the fingerprint generated by the sensor noise pattern. This makes the fingerprint less reliable and, consequently, it should be removed to improve the quality of the noise pattern.

In addition, it must be considered that most of the techniques proposed in the scientific community consider improving the sensor noise fingerprint adapted to digital images Sandoval Orozco et al. (2015). There are few works that adapt the improvement of the fingerprint in the case of digital videos from mobile devices. This last point is the one addressed in the our research work.

Regarding video clustering, this is done in real life scenarios using the standard correlation as a measure of similarity to achieve a correct classification of digital videos by device.

In order to identify the source of digital videos, this proposal has taken into account the effect produced by the high compression ratios that a video contains due to the redundancy that a video contains. To obtain the fingerprint of the sensor noise pattern, the hypothesis proposed by Li (2010) has been considered. Here it is indicated that the images may be severely contaminated by the details of the scene. This proposal has also dealt with the particular case of digital videos. The overall scheme of the proposed technique is shown in Figure 3. The proposed algorithm is divided into 3 main phases that are detailed below:

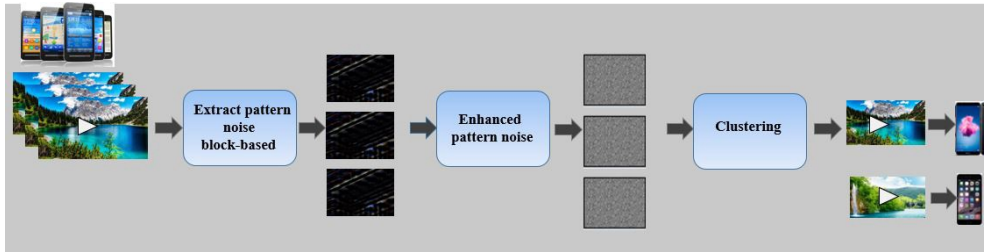

Figure 3: Diagram of the proposed algorithm

1. PRNU noise is stochastic in nature and unique to each sensor. Its high dimensionality and robustness in the processing make it an ideal candidate for forensic applications such as the identification of digital cameras. The extraction of the PRNU noise is done at the macroblocks level as proposed in Kouokam and Dirik (2019). This extraction of noise based on macroblocks takes into account that for each of the frames that make up the video, only those macroblocks are used in which the PRNU noise has not been completely degraded by the compression that a digital video contains are used.

    As it was demonstrated in the work Kouokam and Dirik (2019), the content of a decoded macrblock depends, to a large extent, on the inverse transform DCT of the prediction residual of the block. Therefore the first step that must be performed to To estimate the PRNU fingerprint, is to decode the input video with the version of *jm.*16.1 used in Kouokam and Dirik (2019). The result is a XML file format that contains all the information of the macroblocks that make up the set of frames that belong to a digital video. Figure 4 you can see an extract from the XML file.

    The *Picture* label refers to a frame in the video. The *MacroBlock* tag corresponds to all macroblocks of size 16*x*16 that exist in the frame. The *Position* label indicates the start and end (x, y) coordinates of

```xml
<Picture id="0" poc="0">
  <TypeString>SLICE_TYPE_I</TypeString>
    <MacroBlock num="0">
        <Position>
            <X>0</X>
            <Y>0</Y>
        </Position>
            <PredModeString>BLOCK_TYPE_I</PredModeString>
            <Coeffs>
                <Row>0,0,0,0</Row>
                <Row>0,0,0,0</Row>
                <Row>0,0,0,0</Row>
                <Row>0,0,0,0</Row>
            </Coeffs>
    </MacroBlock>
    . . .
    <MacroBlock num="0">
        <Position>
            <X>752</X>
            <Y>0</Y>
        </Position>
            <PredModeString>BLOCK_TYPE_I</PredModeString>
            <Coeffs>
                <Row>7,6,6,0</Row>
                <Row>6,6,7,6</Row>
                <Row>2,5,6,7</Row>
                <Row>1,0,5,5</Row>
            </Coeffs>
    </MacroBlock>
```

Figure 4: Output file of the decoder stage of a digital video.

the macroblock within the frame. Finally, the *Coeffs* tag indicates the pixel value of the macroblock to be processed.

The macroblocks-based noise extraction used in this work has considered that for each of the frames contained in a video, only those macroblocks are used in which the noise PRNU has not been completely degraded by the compression it contains a digital video.

In consideration of PRNU noise of a macroblock survives compression if the residual coefficients of the macroblock DCT-AC are not all zero, as proposed in Kouokam and Dirik (2019). Only those macroblocks that contain at least a non-zero coefficient DCT-AC will be used in any of the type I, B or P frames that make up the video, otherwise it is discarded the macroblocks. Therefore, a binary matrix will be created with values 0 or 1, whose value is equal to 1 if the macroblocks contains at least a non-null coefficient DCT-AC and 0 in another case. The zeros in the mask mask of the frame indicate the location of pixels or macro bolts where noise estimation PRNU is not feasible. For each of the frames that make up the video, a matrix called mask $M_j$ is obtained. Each of its elements at the location of the pixel ($p; q$) is calculated according to the Equation 1.

$$M_j(p, q) = \begin{cases} 0, & \text{if the DCT-AC coefficients are all zero} \\ 1, & \text{otherwise} \end{cases} \quad (1)$$

Therefore, the fingerprint of the $K$ camera and the video noise of the video $W_j$ are calculated according to the equation 2 where it is obtained from the noise of the video and therefore the fingerprint of the camera.

$$K = \frac{\sum_{j=1}^{n} W_j I_j M_j}{\sum_{j=1}^{n} (I_j M_j)^2 + L} \quad (2)$$

where $I_j$, is the decoded image of the *j-th* frame of the video, $W_j$ is the noise PRNU estimated from $I_j$, $M_j$ is the mask of the frame $j - th$, $n$ is the number of frames of the video to identify and finally, $L$ is an array with all its elements with a value of 1 whose dimension coincides with the resolution of the video to identify. It is necessary to use this matrix, to avoid division by zero in case $M_j(p, q) = 0 \ \forall j$. This matrix called mask is a three-dimensional matrix that includes the dimensions of the input video and another additional dimension, whose value is the number of frames the video has. That is to say, if the video has a dimension of 1920x1080 pixels and a total of 302 frames, then the dimension of the mask array $M_j$ will be 1920x1080x320 elements. This matrix is filled with the number 1 if there is at least a coefficient other than 0 in its positions DCT- AC and with the data 0 in another case.

As an example, in Figure 4 it can be seen that the macroblock number 0 contains the coefficient 0 in all positions DCT-AC of the macroblock. Therefore, the mask values for those 16 elements are all filled with the number 0, while the macroblock number 47 contains more than one non-zero data point in their positions DCT-AC, as visible in the second row of the *Row* tag that contains the coefficients (6, 6, 7, 6). In this case, the values of the mask matrix are filled with the coefficient 1.

This step is computationally expensive due to the large amount of information that the algorithm must process. The calculations that it has made in algorithm with the experiments were carried out on a computer with a $9-th$ generation processor of the Intel Core i7 processor family equipped with 6 cores. Likewise, as can be seen in Figure 4, not all type I frames survive compression. Therefore, it is necessary to use all types of frames (I, B and P) and not only the type I frames, as suggested in other literature proposals.

2. The next step is to improve the PRNU noise fingerprint calculated in the previous step. The details of the scene when representing the high-frequency components make its magnitude much greater than that of the noise pattern. For this reason, it is necessary to remove fragments from the high-frequency scene to improve the sensor noise fingerprint. Subsequently, in Li (2010) an approach is proposed to mitigate the influence of scene details when calculating the noise pattern. In this way Hence, the hit rate is improved when identifying the device on which the proposed algorithm is based. The hypothesis proposed in Li (2010) suggests that an improved fingerprint $K_{enh}$ can be obtained by assigning less significant weighting factors to the strong components of the signal in the domain of the wavelet transform to attenuate the interference of scene details. To test the hypothesis proposed in Li (2010), five models identified as enhancer models (2 - 6) were proposed. Results are detailed in the Experiments section. The enhancer model that obtained the best results in this work was the model called Non-linear Exponential Transformation (Model 3), whose equation is detailed in Equation 3.

$$K_{enh}(i,j) = \begin{cases} 1 - e^{-K(i,j)} & 0 \leq K(i,j) \leq \alpha \\ (1 - e^{-\alpha})e^{\alpha - K(i,j)} & K(i,j) > \alpha \\ -1 + e^{K(i,j)} & -\alpha \leq K(i,j) < 0 \\ (-1 + e^{-\alpha})e^{\alpha + K(i,j)} & K(i,j) < -\alpha \end{cases} \quad (3)$$

where the value of the parameter $\alpha$ determines the attenuation of the noise. In order to decide the value of $\alpha$, various experiments were carried out on videos with different values of $\alpha$ selected at random and that dealt with the majority of possible cases, therefore the following values $\alpha \in [2, 5, 7, 20, 50]$. The value chosen was 20 as it stated to be a prototype value for all the proposed enhancer models and with which desirable results were obtained in the enhancer models used. The output of this step is a two-dimensional matrix containing the improved noises for each of the input videos under investigation. The general scheme of block-based noise extraction together with the enhancement function is shown in Figure 5.

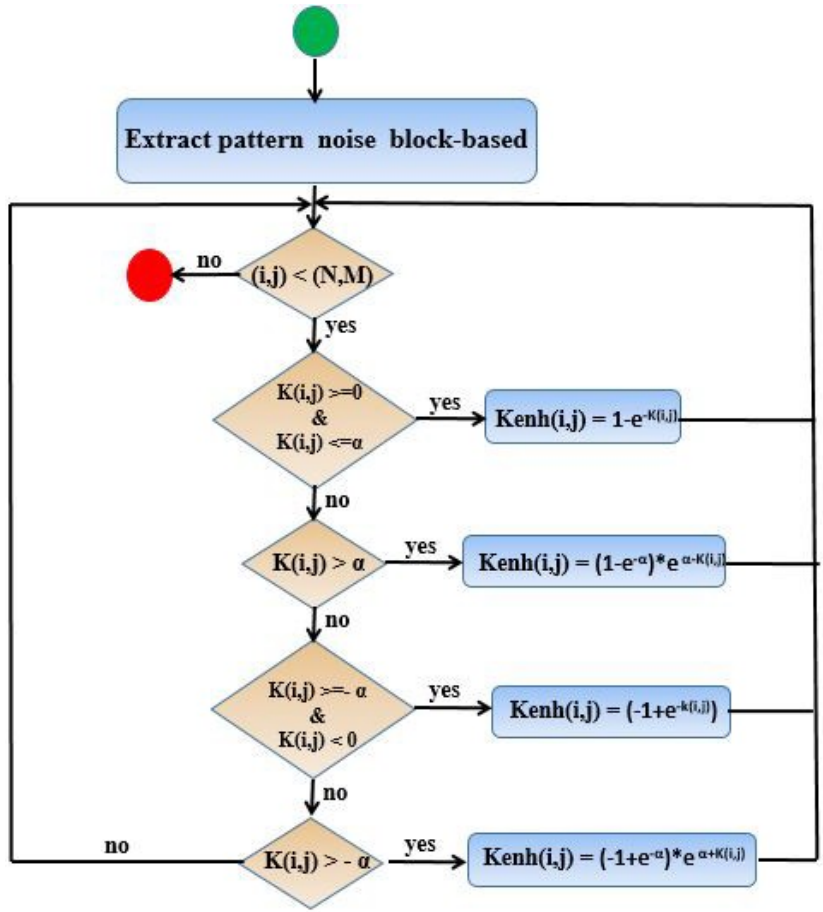

Figure 5: Enhanced algorithm structure.

For each of the improved PRNU noises calculated in the previous step ($K_{enh_1}$, ..., $K_{enh_n}$), the correlation value is obtained using the result of Equation 4.

$$corr(n_i, n_j) = \frac{(n_i - \overline{n_i}) \emptyset (n_j - \overline{n_j})}{||n_i - \overline{n_i}|| \cdot ||n_j - \overline{n_j}||} \quad (4)$$

where $\overline{n_i}$ y $\overline{n_j}$ represent the mean of the vector, $n_i \emptyset n_j$ is the dot product of two vectors, and $||n_i||$ is the norm $L_2$. Since the sensor noise pattern is a two-dimensional matrix, prior to the application of the correlation function, a transformation of the matrix to a one-dimensional vector is performed.

3. Finally, clustering using an unsupervised agglomerative clustering algorithm proposed in some works such as Caldelli et al. (2010), García Villalba et al. (2015). To determine the similarity between videos that belong to the same device, there are distance measures such as: Euclidean distance, Manhattan or Chebychev distance, among others. One of the measures widely used in identifying the source of digital images can be found Caldelli et al. (2010), García Villalba et al. (2015).

To decide how to calculate the correlation value iteratively, it is necessary to define a measure of similarity between the clusters, so that similar clusters are cataloged before different clusters. To carry out this union, a linkage criteria function is performed. This function measures the similarity between two clusters in which at least one of them is made up of more than one sample or video. For this work, tests were carried out with various linkage criteria functions that are simple, average and complete criteria. After conducting several experiments, the linkage criteria function that showed the best results was the average linkage criteria, as in the case of the works related to images Caldelli et al. (2010) and García Villalba et al. (2015). The Equation 5 shows the average linkage criteria function between two groups $u$ and $v$.

$$d(u, v) = \sum_{i \in u, j \in v} \frac{corr(K_{enh_i}, K_{enh_j})}{|u||v|} \quad (5)$$

where, $|u|$ and $|v|$ is the cardinality of the groups of $u$ and $v$ respectively. The value of $corr(K_{enh_1}, K_{enh_n})$ determines the similarity that exists between two objects of the same group, reaching its highest value the more similar $K_{enh_i}$ and $K_{enh_j}$. The value of $corr$ is calculated in the Equation 4.

In traditional hierarchical clustering algorithms, the final result is that all the elements belong to the same group, but in this work we need each group to represent a model at the end of the execution. In order to validate the groups, the silhouette coefficient has been used as a group validation measure. For each observation $x$ there are two measures that are:

- Cohesion $a(x)$: it is obtained as the average distance of $x$ to all the points in the same class.
- Separation $b(x)$: measures the average distance of a point from one of the groups to all other nearby groups. The most used separation is the average distance between $x$ and all the elements in the closest group, although other measures are also used to value the separation.

The coefficient $s(x)$ takes the values in the range [1,1]], where1 corresponds to a bad choice of the number of classes and 1 indicates well defined classes. The silhouette coefficient for $x$ is illustrated in the Equation 6.

$$s_x = \frac{b(x) - a(x)}{max\{a(x), b(x)\}} \qquad (6)$$

and for all the clustering is determined by the Equation 7.

$$SC = \frac{1}{n} \sum_x s(x) \qquad (7)$$

where $n$ is the number of observations.

To assess the quality of the clustering resulting from the clustering algorithm, the precision, the curve ROC and the area AUC are described. The complexity of the algorithm increases as the resolution of the video increases. In the case of a 1920x1080 resolution, a total of 2.73.600 iterations will be necessary, resulting in a high computational cost. The time complexity is $O(n^2)$.

## 4. Experiments and Results

This section details the experiments carried out regarding the identification of the source of digital videos. The first experiment involves finding out which enhancer function of those proposed in Li (2010) is the one that obtains the best results by varying the parameter of $\alpha$. The following experiment is performed on a different data set than the first to prove that the choice of the enhancement model and the parameter $\alpha$ are not linked to a specific video set. This second set of videos used takes into account that the videos have been downloaded from social media platforms such as YouTube or WhatsApp. We must take into account that when a video is uploaded or downloaded from this type of platform, the video undergoes another compression process limiting the quality of the extracted sensor fingerprint.

To evaluate the effectiveness of the smartphone device source identification algorithm, a subset of the public repository ACID Hosler et al. (2019) has been used. In this work, models from smartphone devices have been selected because that is the case study that has been analysed in this paper. Table 1 shows the summary of the video set used.

Table 1: Composition of the ACID video set

| Brand | Model | ID | #Videos | Resolution |
|---|---|---|---|---|
| Huawei | Honor 6X Pixel 2 | M12 | 10 | 1920x1080p |
| LG | X Charge | M17 | 10 | 1920x1080p |
| Samsung | Galaxy J7 Pro | M27 | 10 | 1920x1080p |
| | Galaxy S3 | M28 | 10 | 1920x1080p |
| | Galaxy S5 | M29 | 10 | 1920x1080p |
| | Galaxy S7 | M30 | 10 | 1920x1080p |
| | Galaxy Tab A | M31 | 10 | 1920x1080p |
| | J5-6 | M32 | 10 | 1920x1080p |

### 4.1. Influence of the parameter $\alpha$ on the TPR

The first experiment that has been conducted involves testing all the enhancer models proposed in Li (2010). Table 2 shows the summary of the experiments carried out. It can be seen that when $\alpha = 20$, all the enhancer models ($\gamma_2, \gamma_3, \gamma_4, \gamma_5, \gamma_6$) have the highest success rate, reaching over 81% in almost all the proposed models. The model with the worst average rate is $\gamma_4$, function that never exceeds 25% in TPR.

Table 2: Average TPR by model and with different values of α

| Enhancer Model | Values of α | | | | | TPR Average |
|---|---|---|---|---|---|---|
| | 2 | 5 | 7 | 20 | 50 | |
| $\gamma_2$ | 72% | 70% | 70% | 70% | 60% | 68,4% |
| $\gamma_3$ | 56% | 81,25% | 70% | 81,25% | 81,25% | 73,95% |
| $\gamma_4$ | 25% | 20% | 22 % | 25 % | 20% | 20,44% |
| $\gamma_5$ | 55% | 65% | 56,4% | 81,25% | 55,4% | 62,61% |
| $\gamma_6$ | 60% | 69% | 81,25% | 81,25% | 81,25% | 74,75% |

In Tables 3, 4 and 5 respectively, we can see, in detail, the result of clustering the $\gamma_2$, $\gamma_3$ and $\gamma_6$ functions, since they are the models that have achieved the best average rate when $\alpha = 20$, as can be seen in Table 2.

To calculate the completeness or TPR of each group, it is necessary to see in the group the smartphone model that contains the largest number of videos compared to the total number of videos per model, since that is the predominant model within the group. Next, the percentage of videos that have been correctly classified for that model within a group is calculated. The vast majority of cases can be evaluated as a group associated with one or more devices. In some cases, there are groups in which no device prevails, as can be seen in 3. Its TPR is considered to be 0. Finally, the percentage of videos that have been appropriately classified for that device in that group is calculated.

For the $\gamma_2$ with $\alpha = 20$ function, it can be observed that the groups G8 have $TPR = 0$, because there is no distinction between the M17 and M32 models respectively. It can be seen that the $\gamma_3$ with $\alpha = 20$ function has a higher average TPR than the $\gamma_2$ with $\alpha = 20$ function. No group that has obtained this average by implementing this enhancer has a $TPR = 0$, as was observed in the $\gamma_2$ with $\alpha = 20$ function that had a group with $TPR = 0$. The M12, M27, M28, M29 and M31 devices have the same behavior in both models and they achieve a value of $TPR = 100\%$.

It should be noted that $\gamma_2$ and $\gamma_6$ functions with $\alpha = 20$ value generate the same behavior and the same number of clusters. In order to measure the performance of the clustering proposed in this technique, a curve ROC has been used that shows the productivity of the classification model at all clustering thresholds, in the case of $\gamma_3$ with $\alpha = 20$ function as shown in Figure 6.

Table 3: Clustering result with $γ_2$ with $α = 20$ function

| Brand-Model | G1 | G2 | G3 | G4 | G5 | G6 | G7 | G8 | G9 |
|---|---|---|---|---|---|---|---|---|---|
| M17 | **8** | 0 | 0 | 0 | 0 | 0 | 0 | **2** | 0 |
| M32 | 0 | **5** | 0 | 0 | 0 | 0 | 0 | **2** | **3** |
| M31 | 0 | 0 | **10** | 0 | 0 | 0 | 0 | 0 | 0 |
| M12 | 0 | 0 | 0 | **10** | 0 | 0 | 0 | 0 | 0 |
| M29 | 0 | 0 | 0 | 0 | **10** | 0 | 0 | 0 | 0 |
| M28 | 0 | 0 | 0 | 0 | 0 | **10** | 0 | 0 | 0 |
| M27 | 0 | 0 | 0 | 0 | 0 | 0 | **10** | 0 | 0 |

Table 4: Clustering result with $γ_3$ with $α = 20$ function

| Brand-Model | G1 | G2 | G3 | G4 | G5 | G6 | G7 | G8 | G9 |
|---|---|---|---|---|---|---|---|---|---|
| M32 | **7** | 0 | 0 | 0 | 0 | 0 | 0 | **3** | 0 |
| M27 | 0 | **10** | 0 | 0 | 0 | 0 | 0 | 0 | 0 |
| M31 | 0 | 0 | **10** | 0 | 0 | 0 | 0 | 0 | 0 |
| M28 | 0 | 0 | 0 | **10** | 0 | 0 | 0 | 0 | 0 |
| M17 | 0 | 0 | 0 | 0 | **8** | 0 | 0 | 0 | **2** |
| M29 | 0 | 0 | 0 | 0 | 0 | **10** | 0 | 0 | 0 |
| M12 | 0 | 0 | 0 | 0 | 0 | 0 | **10** | 0 | 0 |

Table 5: Clustering result with $γ_6$ with $α = 20$ function

| Brand-Model | G1 | G2 | G3 | G4 | G5 | G6 | G7 | G8 | G9 |
|---|---|---|---|---|---|---|---|---|---|
| M32 | **7** | 0 | 0 | 0 | 0 | 0 | 0 | **3** | 0 |
| M27 | 0 | **10** | 0 | 0 | 0 | 0 | 0 | 0 | 0 |
| M31 | 0 | 0 | **10** | 0 | 0 | 0 | 0 | 0 | 0 |
| M28 | 0 | 0 | 0 | **10** | 0 | 0 | 0 | 0 | 0 |
| M17 | 0 | 0 | 0 | 0 | **8** | 0 | 0 | 0 | **2** |
| M29 | 0 | 0 | 0 | 0 | 0 | **10** | 0 | 0 | 0 |
| M12 | 0 | 0 | 0 | 0 | 0 | 0 | **10** | 0 | 0 |

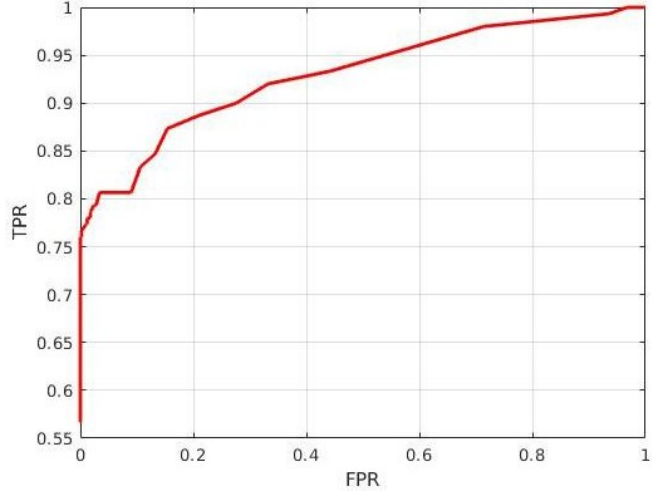

Figure 6: ROC curve for $γ_3$ with $α = 20$ function for ACID video dataset

The value of AUC for the area of the curve ROC in Figure 6 is 0.929980.

*4.2. Comparison between $γ_3$ with $α = 20$ function and block-based method*

In this section we are going to compare the block-based Kouokam and Dirik (2019) proposal where we only look at the effect that compression has when calculating the PRNU fingerprint - with the contribution proposed in this work, where in addition to taking into account the block-based method, the function has been included enhancer that gave the best results in the previous section, that is, $γ_3$ with $α = 20$ function. On the one hand, if we use the videos in Table 1, the same TPR results are obtained, generating the same groups, for the case of the function without improvement (the one proposed in block-based) as the function proposed by this work. The result of both proposals can be seen in Table 4.

On the other hand, if we use of videos in Table 6, where the M00 model, an Apple device, is added, the clusters, seen in Tables 7 and 8, are obtained. In Table 8, it can be observed that, if we do not apply only the block-based method, groups G8 and G9 have a $TPR = 0$, because there is a mix between the M00, M17 and M32 models in group G8. Additionally, it must be taken into account that the Apple brand M00 model cannot be identified with this proposal.

Table 6: Composition of the ACID video set with M00 model is included

| Brand | Model | ID | #Videos | Resolution |
|---|---|---|---|---|
| Apple | iPhone 8 Plus | M00 | 8 | 1920x1080p |
| Huawei | Honor 6X Pixel 2 | M12 | 10 | 1920x1080p |
| LG | X Charge | M17 | 10 | 1920x1080p |
| Samsung | Galaxy J7 Pro | M27 | 10 | 1920x1080p |
| | Galaxy S3 | M28 | 10 | 1920x1080p |
| | Galaxy S5 | M29 | 10 | 1920x1080p |
| | Galaxy S7 | M30 | 10 | 1920x1080p |
| | Galaxy Tab A | M31 | 10 | 1920x1080p |
| | J5-6 | M32 | 10 | 1920x1080p |

Table 7: Clustering result with $\gamma_3$ with $\alpha = 20$ function

| Brand-Model | G1 | G2 | G3 | G4 | G5 | G6 | G7 | G8 | G9 | G10 | Average TPR |
|---|---|---|---|---|---|---|---|---|---|---|---|
| M32 | **7** | 0 | 0 | 0 | 0 | 0 | 0 | **3** | 0 | 0 | |
| M27 | 0 | **10** | 0 | 0 | 0 | 0 | 0 | 0 | 0 | 0 | |
| M31 | 0 | 0 | **10** | 0 | 0 | 0 | 0 | 0 | 0 | 0 | |
| M28 | 0 | 0 | 0 | **10** | 0 | 0 | 0 | 0 | 0 | 0 | |
| M17 | 0 | 0 | 0 | 0 | **8** | 0 | 0 | 0 | 0 | **2** | |
| M29 | 0 | 0 | 0 | 0 | 0 | **10** | 0 | 0 | 0 | 0 | |
| M12 | 0 | 0 | 0 | 0 | 0 | 0 | **10** | 0 | 0 | 0 | |
| M00 | 0 | 0 | 0 | 0 | 0 | 0 | 0 | **3** | **5** | 0 | |
| **TPR-Group** | 70 | 100 | 100 | 100 | 80 | 100 | 100 | 0 | 50 | 20 | 72% |

Table 8: Clustering result with block-based method

| Brand-Model | G1 | G2 | G3 | G4 | G5 | G6 | G7 | G8 | G9 | Average TPR |
|---|---|---|---|---|---|---|---|---|---|---|
| M32 | **5** | 0 | 0 | 0 | 0 | 0 | 0 | **2** | **3** | |
| M27 | 0 | **10** | 0 | 0 | 0 | 0 | 0 | 0 | 0 | |
| M31 | 0 | 0 | **10** | 0 | 0 | 0 | 0 | 0 | 0 | |
| M28 | 0 | 0 | 0 | **10** | 0 | 0 | 0 | 0 | 0 | |
| M17 | 0 | 0 | 0 | 0 | **7** | 0 | 0 | **3** | 0 | |
| M29 | 0 | 0 | 0 | 0 | 0 | **10** | 0 | 0 | 0 | |
| M12 | 0 | 0 | 0 | 0 | 0 | 0 | **10** | 0 | 0 | |
| M00 | 0 | 0 | 0 | 0 | 0 | 0 | 0 | **2** | **6** | |
| **TPR-Group** | 50 | 100 | 100 | 100 | 70 | 100 | 100 | 0 | 0 | 68% |

Therefore, we can conclude that the technique proposed in this contribution improves or equals the average TPR, as compared to only using the block-based method, but does not worsen the TPR obtained with the work done with the block-based method.

*4.3. Average TPR with Videos from Youtube and WhatsApp Social Platforms*

To carry out this experiment, another set of videos available in the scientific community called VISION Shullani et al. (2017) has been used. Unlike the data set used in the previous experiment, this set of videos contains models of mobile devices that have been uploaded to the YouTube and WhatsApp social media platforms, in addition to videos from mobile devices.

A forensic analyst cannot always access the mobile device that generated the multimedia video to identify the source of the video, since most of the videos that show a crime are uploaded on the Internet. To tackle this problem, it is necessary to implement techniques that materialize the identification of videos from social platforms and verify that the fingerprint PRNU remains reliable in this type of situations.

Also in an open stage, it is not very common to have the same number of videos for each device be identified. For this reason, experiments were carried out where the video sets of each of the models have an asymmetric distribution. These experiments were able to determine if the algorithm adapts to real life scenarios.

Table 9 shows the set of videos that has been used for this experiment.

Table 10 summarizes the experimental conditions used in this experiment.

As for the enhancer model to be used, the $\gamma_3$ with $\alpha = 20$ function has been chosen. This model gave us the best results in the previous experiment. As such, when testing this model in another data set, it will be shown that these values do not depend on the specific dataset.

As can be seen in Table 11, the number of videos per device is varied, and yet all the videos have been classified in a single group with the exception of device D03, which has needed 2 groups to be classified correctly.

Figure 7 shows the ROC curve for this dataset.

Table 9: Composition of the VISION dataset sample

| ID | Brand | Resolution | # Videos |
|---|---|---|---|
| D01 | Samsung Galaxy S3 Mini | 1280x720 | 27 |
| D03 | Huawei P9 | 1920x 1080 | 11 |
| D04 | LG D290 | 800x480 | 7 |
| D07 | Lenovo P70A | 1280x720 | 4 |
| D09 | Apple iPhone 4 | 1280x720 | 3 |
| D13 | Apple iPad2 | 1280x720 | 3 |
| D16 | Huawei P9Lite | 1920x1080 | 3 |
| D17 | Microsoft Lumia 640 | 1920x1080 | 3 |
| D21 | Wiko Ridge 4G | 1920x1080 | 3 |
| D22 | Samsung Galaxy Trend Plus | 1280x720 | 3 |
| D24 | Xiaomi_RedmiNote3 | 1920x1080 | 3 |
| D25 | OnePlus A3000 | 1920x1080 | 3 |
| D33 | Huawei Ascend | 1280x720 | 3 |

Table 10: Experimental conditions

| Parameter | Value |
|---|---|
| Minimum number of videos from Social Platforms (YouTube, WhatsApp) | 3 |
| Enhancer model | $\gamma_3$ |
| Value of $\alpha$ | 20 |

The value of the curve AUC for the area of the curve ROC in Figure 6 is 0.956802.

As seen in the experiments section, the highest success rate is achieved with the improvement model $\gamma_3$ with $\alpha = 20$. It is demonstrated that it is necessary to apply techniques that improve the PRNU fingerprint of the sensor in the case of digital videos to improve the hit rate.

Table 11: TPR with devices from social platforms

| ID | G1 | G2 | G3 | G4 | G5 | G6 | G7 | G8 | G9 | G10 | G11 | G12 | G13 | G14 |
|---|---|---|---|---|---|---|---|---|---|---|---|---|---|---|
| D01 | **27** | 0 | 0 | 0 | 0 | 0 | 0 | 0 | 0 | 0 | 0 | 0 | 0 | 0 |
| D03 | 0 | **2** | **9** | 0 | 0 | 0 | 0 | 0 | 0 | 0 | 0 | 0 | 0 | 0 |
| D04 | 0 | 0 | 0 | **7** | 0 | 0 | 0 | 0 | 0 | 0 | 0 | 0 | 0 | 0 |
| D07 | 0 | 0 | 0 | 0 | **4** | 0 | 0 | 0 | 0 | 0 | 0 | 0 | 0 | 0 |
| D09 | 0 | 0 | 0 | 0 | 0 | **3** | 0 | 0 | 0 | 0 | 0 | 0 | 0 | 0 |
| D13 | 0 | 0 | 0 | 0 | 0 | 0 | **3** | 0 | 0 | 0 | 0 | 0 | 0 | 0 |
| D16 | 0 | 0 | 0 | 0 | 0 | 0 | 0 | **3** | 0 | 0 | 0 | 0 | 0 | 0 |
| D17 | 0 | 0 | 0 | 0 | 0 | 0 | 0 | 0 | **3** | 0 | 0 | 0 | 0 | 0 |
| D21 | 0 | 0 | 0 | 0 | 0 | 0 | 0 | 0 | 0 | **3** | 0 | 0 | 0 | 0 |
| D22 | 0 | 0 | 0 | 0 | 0 | 0 | 0 | 0 | 0 | 0 | **3** | 0 | 0 | 0 |
| D24 | 0 | 0 | 0 | 0 | 0 | 0 | 0 | 0 | 0 | 0 | 0 | **3** | 0 | 0 |
| D25 | 0 | 0 | 0 | 0 | 0 | 0 | 0 | 0 | 0 | 0 | 0 | 0 | **3** | 0 |
| D33 | 0 | 0 | 0 | 0 | 0 | 0 | 0 | 0 | 0 | 0 | 0 | 0 | 0 | **3** |

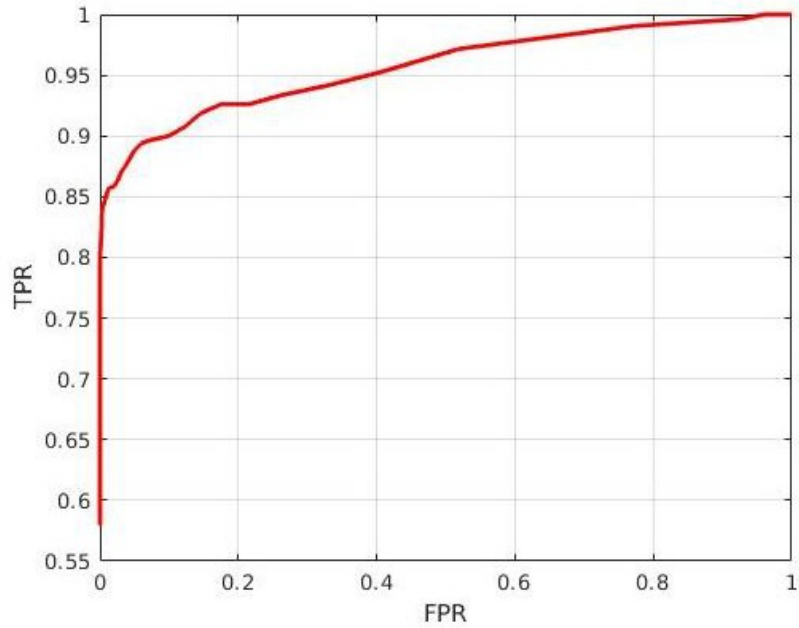

Figure 7: ROC curve the $\gamma_3$ with $\alpha = 20$ function for VISION video dataset

## 5. Conclusions and Future Work

The general conclusion is that the presented technique is valid and achieves good results. The algorithm that extracts the sensor fingerprint takes into account the effect of video compression as well as the removal of high-frequency details contained in the scenes of the frames to improve the hit rate. It should be noted that in the most recent literature, techniques can be found that improve the fingerprint pattern of sensor imperfections applied to digital images but not to digital videos. This work shows that it is necessary to improve the sensor footprint also in the case of digital videos.

The proposed clustering algorithm is based on the combination of a hierarchical and a flat clustering. Most of the groups that have been obtained are pure, that is, they are made up o a single mobile device. The percentage of success when using a smaller clipping size reduces the hit rate. But on the other hand, the calculation time is smaller, so a balance should be found between the clipping size and the calculation time.

The unsupervised grouping technique used is carried out on a wide variety of videos, that is, videos from mobile devices (native videos) and videos downloaded from the main video- sharing platforms such as WhatsApp and YouTube. Results higher than 90% in the hit rate were achieved. Therefore, this technique can be used in real scenarios because the analyst does not need to have access to the mobile device that generated the video since the technique also works with videos downloaded from the main multimedia video sharing platforms such as WhatsApp and YouTube.

As future work, the most recent works in the literature dealing with digital images such as Gupta and Tiwari (2018) and Kang et al. (2011) could be applied to videos to see if better results are achieved than those obtained with the Li (2010) technique. Finally, another future line of work that is proposed is associated with the high computational cost of the proposed technique. It is necessary to use metamodels or surrogate models that reduce this computational complexity and help increase the number of videos used in the experiments such as Alizadeh et al. (2019) and Alizadeh et al. (2020).

**Acknowledgements**

This work is promoted under the specialty of Industrial PhD in collaboration with Securitas Direct. This work has also received funding from THEIA (Techniques for Integrity and Authentication of Multimedia Files of Mobile Devices) UCM project (FEI-EU-19-04).